\documentclass{aa}
\usepackage{graphics}

\def\etal{{\rm et al.\/} }

\def\secondip{\hbox{\rlap{\hbox{.}}\hbox{$''$}}}
\def\primip{\hbox{\rlap{\hbox{.}}\hbox{$'$}}}
\def\gradip{\hbox{\rlap{\hbox{.}}\raise 5.truept \hbox{{\small $\circ$}}}}
\begin{document}

   \thesaurus{08     
	       08.16.3;  
	       10.07.3;  
	       10.07.2;  
               08.12.2;  
	       08.12.3;  
)} 

\title{The Luminosity and Mass Function of the Globular 
Cluster NGC~1261
\thanks{Based on observations collected at the European 
Southern Observatory, La Silla, Chile.}}
\author{M. Zoccali\inst{1} \and G. Piotto\inst{1} \and S.R. Zaggia\inst{2}
\and M. Capaccioli\inst{2,3}}
\offprints{M. Zoccali, email: zoccali@astrpd.pd.astro.it}
\institute{Dipartimento di Astronomia, Universit\`a di Padova, vicolo 
dell'Osservatorio 5, I-35122 -- Padova -- Italy 
\and
Osservatorio Astronomico di Capodimonte, via Moiariello 16,
I-80131 -- Napoli -- Italy
\and Dipartimento di Scienze Fisiche, Universit\`a Federico~II, 
Mostra d'Oltremare, Padiglione 19, I-80125 Napoli, Italy.}

\date{Received September 1997 ; accepted 24 November 1997 }

\authorrunning{Zoccali \etal }
\titlerunning{The L.F. and M.F. of NGC~1261}
\maketitle


\begin{abstract}
I-band CCD images of two large regions of the Galactic globular
cluster NGC 1261 have been used to construct stellar luminosity
functions (LF) for 14000 stars in three annuli from $1\primip4$ from
the cluster center to the tidal radius. The LFs extend to $M_I\sim8$
and tend to steepen from the inner to the outer annulus, in agreement
with the predictions of the multimass King--Michie model that we have
calculated for this cluster. The LFs have been transformed into mass
functions. Once corrected for mass segregation the global mass
function of NGC 1261 has a slope $x_0=0.8\pm0.5$
\end{abstract}


\section{Introduction}

One of the basic ingredients of a stellar system is the mass
distribution of its population of stars.  In the case of globular
clusters (GCs), stellar mass functions (MFs) reflect both the initial
conditions from the epoch of cluster formation and the dynamical
evolution.  In order to try to disentangle the effects of evolution
from those of the initial conditions on the shapes of globular cluster
MFs we have started a long-term project aimed at getting deep
luminosity functions (LF) in a large sample of galactic GCs.  The
first results of this project have already been presented in other
papers ({\it cf.} Piotto 1993 and Piotto \etal (1996) for a complete
bibliography on recent works on GC LFs).  Since Scalo (1986) it has
been known that the MFs of Galactic GCs differ from cluster to
cluster. McClure \etal (1986) proposed that the MF slopes depend on
the cluster metallicity. By studying a sample of 17 Galactic GCs,
Capaccioli, Ortolani, \& Piotto (1991) questioned the hypothesis that
the metal content could be the main parameter governing the MF shapes;
instead, they showed that the position of the parent cluster within
the Galactic gravitational potential is affecting the present day MFs
(PDMF).  By means of a more detailed statistical analysis on the same
sample of Capaccioli \etal (1991), Djorgovski, Piotto, \& Capaccioli
(1993) demonstrated that the MF slopes are determined both by the
positional variables and by the metallicity.  The two positional
variables are dominant, with similar but not equivalent effects: at a
given Galactocentric distance $R_{GC}$, clusters with a smaller
distance from the galactic plane $Z_{GP}$ have shallower MFs, and
vice-versa.  However, they do not determine the MF completely: at a
given position, clusters with higher metallicity have shallower MFs.
This dependence on the positional coordinates has been tentatively
interpreted as the effect of selective evaporation of stars, driven by
tidal shocks (Capaccioli, Piotto, \& Stiavelli 1993): heating due to
disk shocking acts preferentially on stars located in the outer
regions of the cluster and, in combination with mass segregation, it
leads to a loss of low mass stars, i.e. to a MF flattening.  The
dependence on the metallicity could reflect the initial conditions.
This scenario has been furtherly confirmed by recent HST data (Piotto,
Cool, \& King 1997) which allowed to extend the study of GC MFs almost
down to the limit of the core hydrogen-burning ignition ($\sim
0.1M_\odot$).

\begin{figure*}
\resizebox{12cm}{!}{\includegraphics{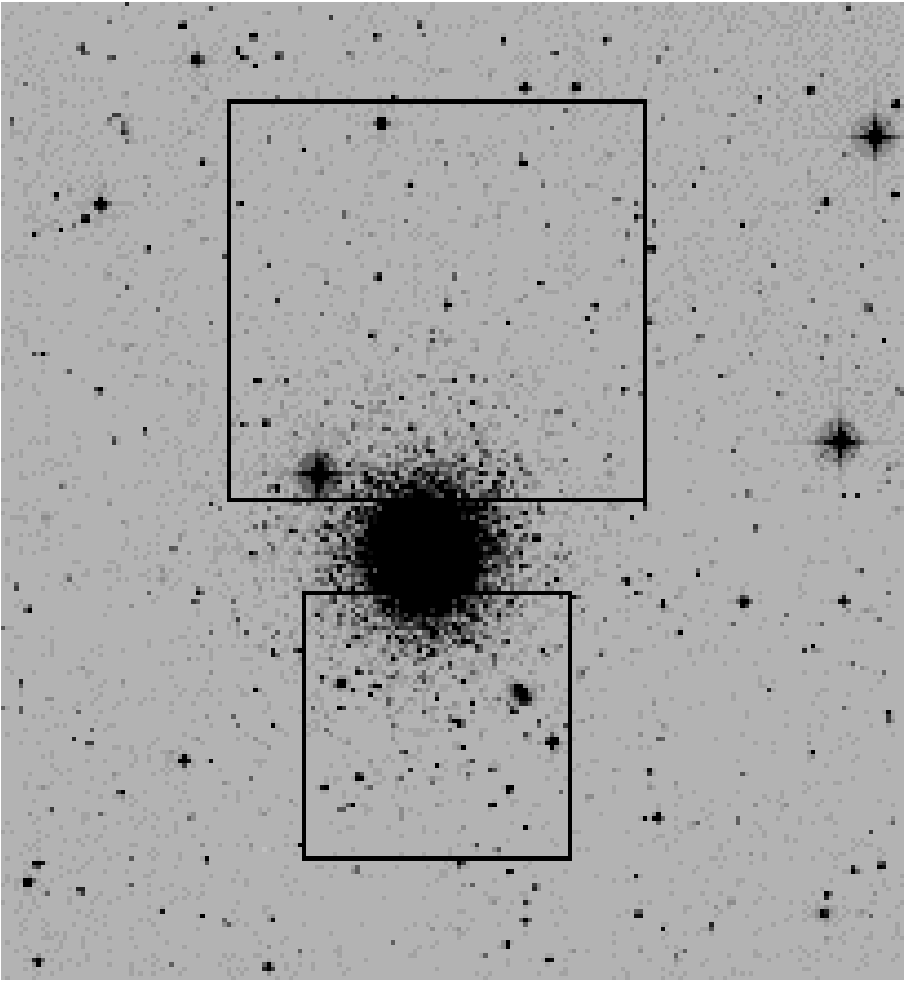}} 
\hfill
\parbox[b]{55mm}{
\caption{The two fields observed in NGC~1261: the upper one is the 
NTT field, while the lower box shows the 2.2m field}
\label{map}}
\end{figure*}

In this paper we present I-band stellar luminosity functions of
NGC~$1261=$C$0310-554$, a cluster which is presently relatively far
from the center and the plane of the Galaxy.  On the basis of the
location of NGC~1261 within the Galaxy, $R_{GC}=17.8$~kpc and
$Z_{GP}=12.6$~kpc, it is very likely that its evolution has been
driven more by the internal dynamics rather than by strong tidal
shocks.  For this reason, despite its large distance from the Sun,
NGC~1261 has been chosen (among other clusters) for a detailed study
of the PDMF, which should be as close as the internal dynamical
evolution can allow to the initial MF (IMF).

NGC~1261 ($\alpha_{2000}=3^h 12^m 15^s$,
$\delta_{2000}=-55^\circ13^\prime1^{''}$; is a medium concentration
globular cluster ($c=1.27$ Trager, King and Djorgovski 1995), located
at high galactic latitude, $l=250\gradip54$, $b=-52\gradip13$.  Its
low reddening and small field star contamination make it a very good
target for photometric studies.  The first CCD color--magnitude
diagram (CMD) for NGC~1261 has been published by Bolte \& Marleau
(1989). They observed the cluster in $B$ and $V$, down to magnitude
$V\sim23$.  From the apparent magnitude of the HB and the fitting of
the main sequence to two subdwarfs, they derived an apparent distance
modulus $(m-M)_V=16.05\pm0.25$.  By matching their observations to the
VandenBerg and Bell (1985) isochrones and assuming E(B-V)=0.02, Bolte
and Marleau obtained an age of 15 Gyr and a metallicity
[M/H]=--1.09. By adopting E(B--V)=0.04, the best fitting isochrone
gives an age of 15 Gyr and a metallicity [M/H]=--1.27.

Alcaino \etal (1992) published a $BVR$ photometric study of NGC~1261.
They found a distance modulus $(m-M)_V=16.00\pm 0.23$, from the
apparent magnitude of the HB, and, using the same set of isochrones of
the previous study, an age of $15\pm2$ Gyr, and a reddening of
E(B-V)$=0.07$, assuming a metallicity [Fe/H]$=-1.27$.  More recently,
Ferraro \etal (1993) presented a $BV$ CMD and a luminosity function
for the evolved stars in this cluster.  From the CMD metallicity
indicators, calibrated with the Zinn \& West (1984) metallicity scale,
they estimated a slightly lower metal content than in the previous
studies: [Fe/H]$=-1.4\pm0.2$.  They also detected for the first time
the `RGB bump' of the cluster, as a clump of stars in the LF of the
red giant branch, at magnitude $V=16.70\pm0.05$.

The paper is organized as follows: in Section 2 we discuss our data
and the method we followed to obtain the photometry. In Section 3 we
present the luminosity functions in three radial annuli.  Section 4
concerns the mass functions and the correction for the effects of the
mass segregation. Finally, in the last Section the MF of NGC~1261 is
compared with the MFs for other 19 Galactic GCs.

\begin{figure}
\resizebox{\hsize}{!}{\includegraphics{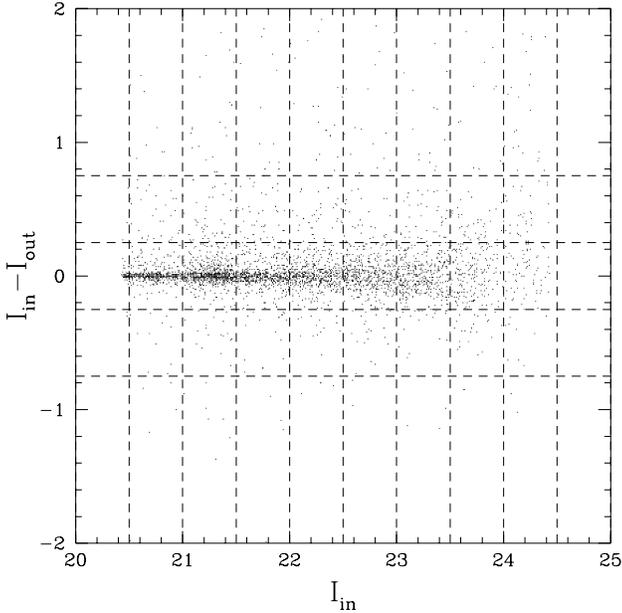}} 
\caption{This plot shows the difference between the magnitude at which
each artificial star was added to the image and the magnitude at which it
was found, as a function of the input magnitude. The graphic refers to the
artificial-star experiments on the NTT frame. The horizontal and vertical
dashed lines are the limits of the LF magnitude bin (i.e. a star that
falls out from the two horizontal lines at $\pm0.250$ would have been
counted in the previous or in the next LF magnitude bin.}
\label{matrix}
\end{figure}


\section{Observation and data reduction}
\label{sec1}

The database consists of two I--band image sets.  The first one is a
sample of seven 15 min exposures obtained on November 27--30, 1991,
with EFOSC2 + CCD \# 17 at the ESO/MPI 2.2~m telescope of La
Silla. These images have dimensions of $1024 \times 1024$ pixels,
corresponding to an area of 5$\primip$8$\times$5$\primip$8 on the sky,
and cover a region from $\sim 1 \primip 4$ to $\sim 7 \primip 0$ from
the center of the cluster.  The second set consists of 9 frames, for a
total exposure time of 95 minutes, obtained with EMMI + CCD \# 34 at
the ESO/NTT telescope in December 1993.  This field cover $9 \primip 6
\times 8 \primip 5$ of the sky, from $\sim 1 \primip 2$ to $\sim 9
\primip 5$ from the center, as shown in Fig.~\ref{map}.
All observations have been taken in fairly good seeing conditions 
($<FWHM> \sim 1 \secondip 0$).

\begin{figure}
\resizebox{\hsize}{!}{\includegraphics{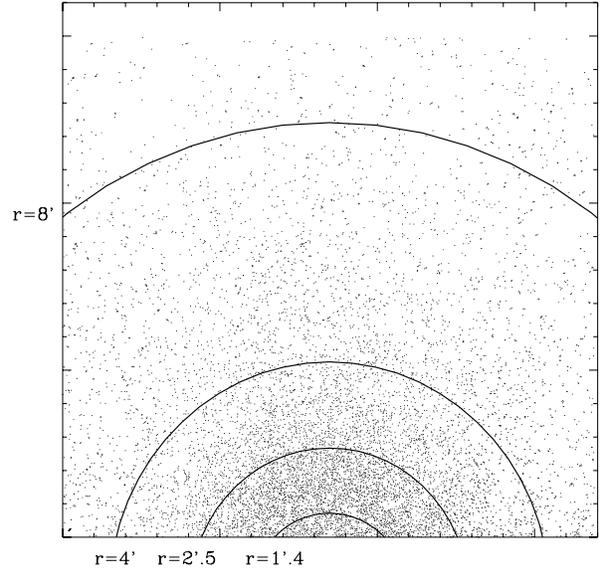}} 
\caption{The position of the stars present in the field observed with
EMMI and the radial annuli in which we separated the counts. The inner
annulus has boundaries $1\primip4\leq r < 2\primip5$, the intermediate 
one has $2\primip 5\leq r < 4\primip 0$, and the outer annulus has 
$4\primip 0\leq r < 8\primip 0$}
\label{field_ntt}
\end{figure}

\begin{figure}
\resizebox{\hsize}{!}{\includegraphics{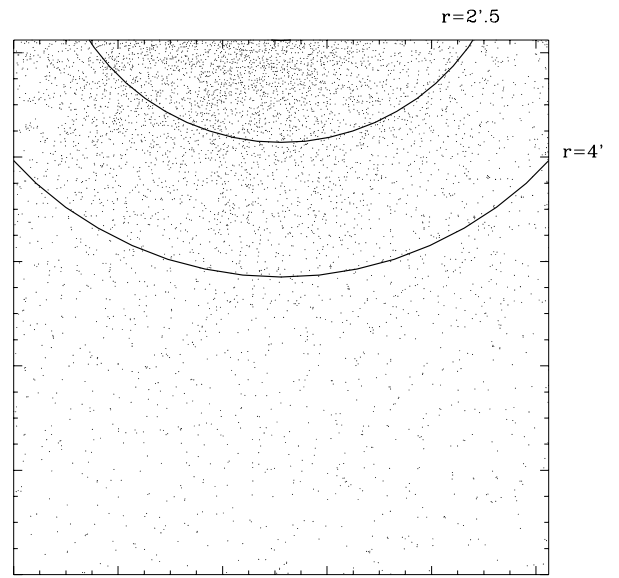}} 
\caption{Same as Fig.~\ref{field_ntt}, for the field observed with EFOSC2}
\label{field_22m}
\end{figure}

The images were bias-corrected, trimmed and flat-fielded with routines 
in MIDAS. The single frames of the same field were averaged in order to 
increase the signal-to-noise ratio. Profile-fitting photometry of stars
was performed with DAOPHOT II (Stetson 1987).

In order to obtain deep luminosity functions it is necessary to
measure the magnitudes of very faint stars with good accuracy.  This
can be dangerous if one has frames only in one photometric band,
because in this case it is very difficult to identify and reject
spurious detections.  For EMMI images we minimized this problem by
averaging separately two sets of frames: a first set of five images
(say image A) and a second one of four images (image B).  The stellar
photometry on these two images has been done independently from each
other.  The star list used as input for the photometry on images A and
B was the final list of objects resulting from the complete photometry
of the median image of all the (nine) original frames. This is the
deepest image, as it has the highest signal-to-noise ratio. The final
star catalog (used for the LFs) contains only the stars measured in
both images A and B. The magnitude associated to each star is the
average of the magnitudes we measured in the two images.
%
%
This method allows to use the full information present in our data set
and also to eliminate most of the spurious detections, as, in general,
the noise peaks cannot be present in both images at the same position.
Indeed, the probability that a pixel differs by more than 3.5 $\sigma$
from the average value of the sky is 0.0004; the probability that the
$same$ pixel differs more than 3.5 $\sigma$ in both images is
$1.6\times 10^{-7}$. Since our images have dimension of
1700$\times$1500 pixels, for a total of 2550000 pixels, we expect that
0.41 noise peaks should be present as spurious detections in our star
list.

The completeness corrections have been estimated by standard
artificial-star experiments. We performed 10 independent experiments
per field, adding each time a total of 600 stars, with the spatial
distribution of a King model of the same concentration (c=1.3, Trager
\etal 1995) of NGC~1261. The finding algorithm adopted to recover and
measure the artificial stars was the same used for the photometry 
of the original images.
As discussed extensively by Stetson \& Harris (1988) and Drukier \etal (1988), 
photometric errors cause the measured magnitude of an artificial star
to be different from its input magnitude, just as ``real'' stars may have
measured magnitudes different from their true magnitudes.

\begin{table}
\caption[]{NTT Luminosity Functions}
\label{tab1}
\[
\begin{array}{lllllll} 
\hline
\noalign{\smallskip}
 I Mag & N_{inn} & N^c_{inn} & N_{int} & N^c_{int} & N_{out} & N^c_{out} \\  
\noalign{\smallskip}
\hline
\noalign{\smallskip}
 16.75   & 2.9    & 2.9\pm0.8     & 0.4   & 0.4\pm0.2   &  -    &  -          \\
 17.25   & 2.7    & 2.6\pm0.8     & 0.5   & 0.4\pm0.2   &  -    &  -          \\ 
 17.75   & 2.9    & 2.9\pm0.8     & 1.0   & 0.9\pm0.3   &  0.2  & 0.1\pm0.1   \\ 
 18.25   & 5.6    & 5.5\pm1.1     & 1.2   & 1.1\pm0.3   &  0.4  & 0.3\pm0.1   \\ 
 18.75   & 6.7    & 6.6\pm1.2     & 2.2   & 2.1\pm0.4   &  0.4  & 0.3\pm0.1   \\ 
 19.25   & 22.2   & 22.0\pm2.2    & 3.6   & 3.5\pm0.5   &  0.7  & 0.6\pm0.1   \\ 
 19.75   & 47.9   & 46.9\pm3.2    & 7.1   & 7.0\pm0.8   &  0.9  & 0.8\pm0.1   \\ 
 20.25   & 62.9   & 62.8\pm3.7    & 11.2  & 11.0\pm0.9  &  2.1  & 2.0\pm0.2   \\ 
 20.75   & 86.2   & 126.7\pm12.5  & 15.2  & 15.0\pm1.4  &  3.3  & 3.1\pm0.3   \\ 
 21.25   & 94.1   & 94.8\pm17.0   & 19.6  & 20.1\pm1.6  &  4.8  & 4.2\pm0.4   \\ 
 21.75   & 96.3   & 136.7\pm25.6  & 27.7  & 30.1\pm2.3  &  6.4  & 5.9\pm0.5   \\ 
 22.25   & 74.1   & 164.9\pm31.1  & 31.4  & 29.3\pm3.1  &  8.9  & 8.1\pm0.6   \\ 
 22.75   & -      & -             & 31.8  & 42.7\pm3.5  & 10.5  & 9.4\pm0.8   \\
 23.25   & -      & -             & 30.1  & 43.8\pm5.9  & 11.5  & 11.5\pm1.1  \\   
 23.75   & -      & -             &  -    &  -          &  8.6  & 12.1\pm5.0  \\   
\noalign{\smallskip}
\hline 
\end{array}
\] 
\end{table}
 
\begin{table}
\caption[]{2.2m Luminosity Functions}
\label{tab2}
\[
\begin{array}{lllllll} 
\hline
\noalign{\smallskip}
I Mag & N_{inn} & N^c_{inn} & N_{int} & N^c_{int} & N_{out} & N^c_{out}  \\  
\noalign{\smallskip}
\hline
\noalign{\smallskip}
 16.75   & 0.9  & 0.9\pm0.5   &  0.6 & 0.6\pm0.3   &   -  &  -          \\
 17.25   & 2.5  & 2.4\pm0.8   &  0.5 & 0.4\pm0.2   &   -  &  -          \\ 
 17.75   & 2.5  & 2.4\pm0.8   &  0.8 & 0.8\pm0.3   &  0.1 & 0.1\pm0.1   \\ 
 18.25   & 2.7  & 2.6\pm0.8   &  1.8 & 1.7\pm0.5   &  0.2 & 0.1\pm0.1   \\ 
 18.75   & 11.1 & 9.2\pm2.7   &  2.8 & 2.8\pm0.6   &  0.6 & 0.5\pm0.2   \\ 
 19.25   & 24.3 & 20.6\pm9.9  &  7.6 & 7.0\pm1.0   &  1.4 & 1.3\pm0.3   \\ 
 19.75   & 37.2 & 33.9\pm18.0 & 13.6 & 11.4\pm1.4  &  2.1 & 1.9\pm0.3   \\ 
 20.25   & 55.1 & 50.7\pm27.5 & 20.8 & 20.7\pm1.9  &  2.6 & 2.5\pm0.4   \\ 
 20.75   & 61.5 & 59.9\pm20.3 & 27.0 & 24.7\pm2.2  &  4.3 & 4.1\pm0.5   \\ 
 21.25   & 68.9 & 94.4\pm27.5 & 35.6 & 33.6\pm3.6  &  4.0 & 3.3\pm0.8   \\ 
 21.75   & 54.0 & 71.7\pm24.5 & 39.6 & 43.3\pm7.1  &  6.3 & 4.6\pm1.3   \\ 
 22.25   & 52.8 & 96.0\pm25.5 & 36.1 & 44.8\pm19.3 &  8.0 & 8.8\pm2.3   \\ 
 22.75   &  -   & -	      & 28.8 & 47.4\pm22.6 &  8.0 & 9.6\pm3.1   \\
 23.25   &  -   & -	      &  -   &    -        &  6.4 & 11.0\pm5.9  \\   
 23.75   &  -   & -           &  -   &    -	   &   -  & -           \\  
\noalign{\smallskip}
\hline
\end{array} 
\]
\end{table}

In general, the measured magnitude of a star is brighter than the
input magnitude, as shown in Fig.~\ref{matrix} (see Stetson \& Harris
1988 for a discussion on the origin of this phenomenon).  A solution
to this problem was suggested by Drukier \etal (1988): using the
artificial--star data it is possible to set up a two dimensional
matrix giving, for each input--magnitude bin, the probability that a
star would appear in each output--magnitude bin.  To be more explicit,
the columns of the matrix represent the input magnitude bins and the
rows represent the output. In this way, each matrix element contains
the probability that a star added in the magnitude bin $i$ (row $i$)
is found in the magnitude bin $j$ (column $j$).  This probability is
defined as the ratio of the number of artificial stars found in each
magnitude bin (rows of the matrix) to the number of stars added in a
bin (columns of the matrix).  The inverse of this matrix gives the
completeness correction factors.  The observed LF, once multiplied by
this inverse matrix, becomes the complete LF, with the two different
effects of crowding properly corrected: the loss of stars and their
migration in magnitude ({\it cf.} Drukier \etal 1988 for the details).
In order to evaluate the uncertainty associated to each element of the
completeness matrix, we performed nine independent experiments on each
field, created nine matrices and inverted each of them separately,
using the Gauss-Jordan elimination method. By calculating the mean
value and its standard deviation for each element, we were able to
determine the errors associated to the crowding correction.

\begin{figure*}
\resizebox{\hsize}{!}{\includegraphics{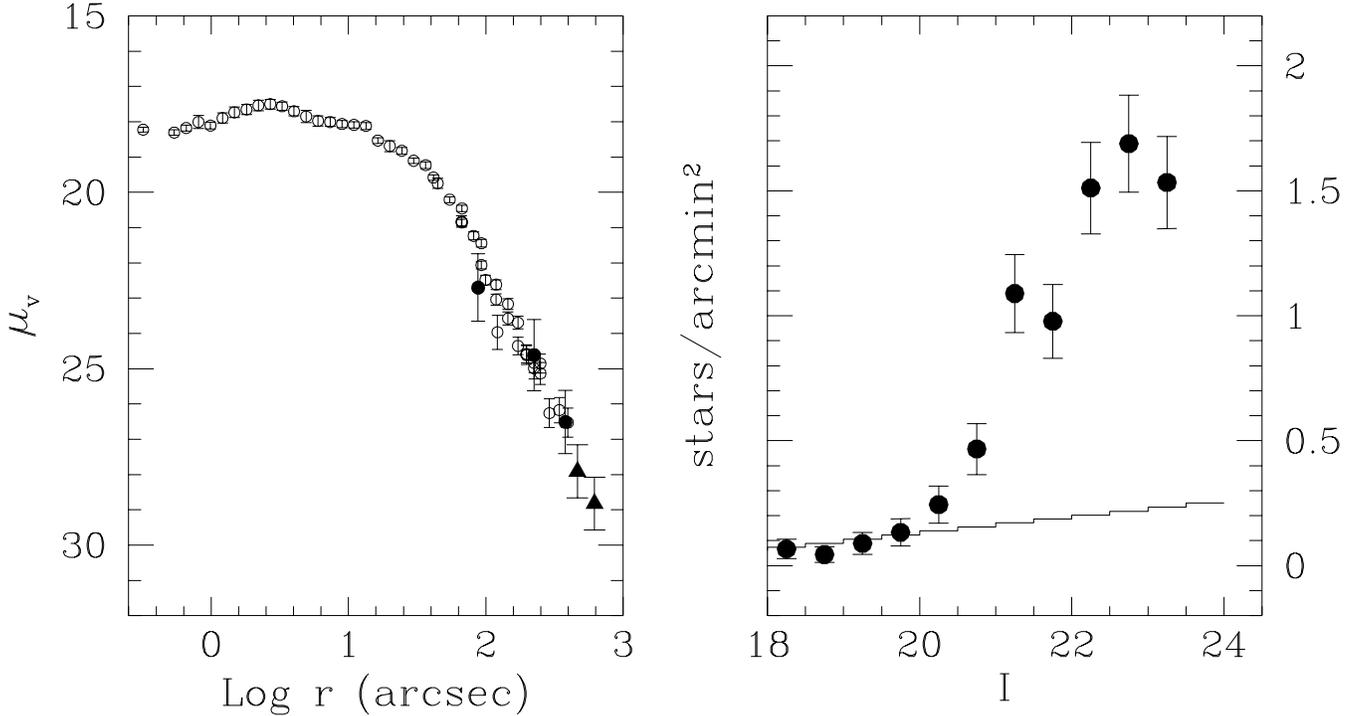}} 
\caption{{\bf Left panel}: {\it open circles} the radial surface 
brightness profile of NGC~1261 by Trager \etal (1995). The points 
relative to our counts in the EMMI field (after completeness correction) 
are represented with {\it filled circles}.  The fourth annulus (which is 
entirely outside the tidal radius given by Trager \etal) has
been divided into two sub-annuli.  
Here we note that the star counts monotonically decrease till 
the outer limit of our field, i.e. we never reach the background level, 
even if the last two points refer to stars which are all outside the tidal 
radius $r_t=7\primip3$ given by Trager \etal {\bf Right panel}: 
comparison between the LF for the annulus outside the published tidal 
radius (filled circles) corresponding to the last two points in the left 
panel, and the background/foreground counts predicted by Ratnatunga \& 
Bahcall (1985) for this region. The observed counts in the right
panel are not corrected for completeness.}
\label{profilo}
\end{figure*}

In order to calibrate the instrumental magnitudes to a standard
system, during each observing night we observed 27 standard stars at
the 2.2m and 18 standards at the NTT, in 4 Landolt (1992) standard
fields.  The calibration equations are:
\begin{displaymath}
I_{instr}-I=(0.04\pm 0.01)(V-I)+2.14\pm 0.01
\end{displaymath}
for the 2.2m data, and:
\begin{displaymath}
I_{instr}-I=(0.004\pm 0.002)(V-I)+0.60\pm 0.01
\end{displaymath}
for the NTT data.  As we have no $V$-band image, we adopted a mean
color $(V-I)=1.2$ for the main-sequence stars in the range of
magnitudes we were studying (17$<M_I<$24).  Since the total range in
color spanned by this region of the main-sequence is
$\Delta(V-I)$=0.8, the real color of our stars might be wrong by at
most $\pm$0.4 magnitudes.  Using the color term of the two equations
above, this gives an error of $\Delta I=\pm 0.02$ for the 2.2m data,
and $\Delta I=\pm 0.002$ for NTT. Considering also the error in the
zero point, we have a total uncertainty $\Delta I=\pm0.03$ for the
2.2m data and $\Delta I=\pm 0.01$ for the NTT data, negligible in both
cases, in view of the fact that the LF bin is of 0.5 magnitudes.


\section{Luminosity Functions}
\label{sec2}

The luminosity function of a system is constructed by simply counting the
number of stars in successive magnitude bins. We separated observed star
counts in different radial annuli (see Figs.~\ref{field_ntt} and
\ref{field_22m}) in order to take into account the different effects of 
crowding at different distances from the center of the cluster. Moreover, 
by determining luminosity and mass functions at different radial distances 
we are able to study the effects of mass segregation (Section V).

For both the EMMI and the 2.2~m fields we separated the star counts in
three annuli: the inner one, from $1 \primip 4$ to $2 \primip 5$, the
intermediate one, from $2 \primip 5$ to $4^\prime$, and the external
one, from $4^\prime$ to $8^\prime$.  The stars in a fourth region of
the NTT image, with $r>8^\prime$, have been used to check the
background/foreground contamination.  The dimensions of these annuli
were chosen in order to have about the same statistical sample of
stars and to avoid a significant crowding gradient inside each one.
The latter condition ensured that we could apply just one completeness
correction for each radial luminosity function.

A peculiarity of NGC~1261 is the fact that it lies on the line of
sight to a cluster of galaxies. These galaxies were identified by
DAOPHOT as agglomerates of stars. To solve this problem, we visually
inspected the images and found the coordinates and the sizes of the
galaxies directly on the frame, and then rejected all the stars within
the galaxy boxes from the output file of ALLSTAR.  

\begin{figure}
\resizebox{\hsize}{!}{\includegraphics{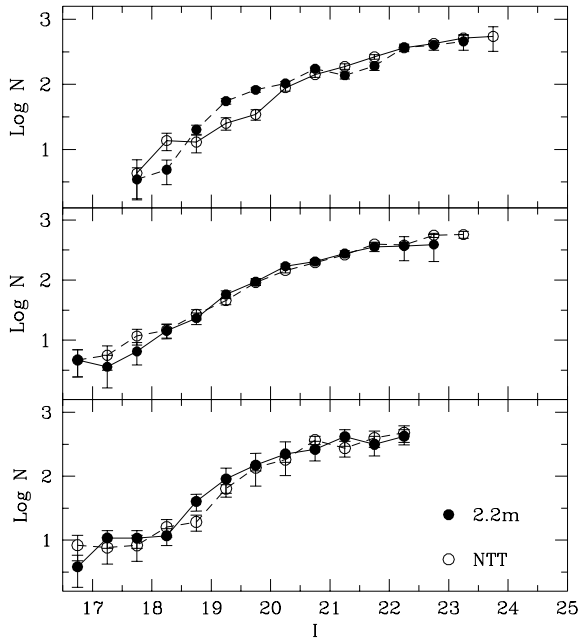}}
\caption{Comparison between the LFs of the two fields, for each radial 
annulus. Bottom to top: inner, intermediate and outer annulus.}
\label{fdl_rad}
\end{figure}

Another source of contamination is given by the foreground/background
stars that are projected on the cluster. Normally one can correct this
effect by determining the luminosity function of a region of sky near
the cluster (in order to have the same population of field stars) but
far enough to be sure there is no cluster object.  Then, one has
simply to subtract the number of field stars per unit area and unit
magnitude from the cluster LF.  Our idea was to consider as a
foreground field the region just outside the tidal radius of NGC~1261,
in the EMMI field.  As the published value of the tidal radius of
NGC~1261 is $r_t=7^\prime.3$ (Trager \etal 1995), we isolated the
stars outside $8^\prime$ from the center of the cluster, and
determined their luminosity function. Unfortunately, a plot of the
cumulative number of stars per unit area versus the distance from the
center demonstrates that there is no flattening at the tidal radius,
i.e. the value of $r_t$ seems to be underestimated (or, alternatively,
there is a halo of cluster stars, beyond the tidal radius, as found by
Grillmair \etal 1995 for other clusters).  This fact is confirmed by
Fig.~\ref{profilo} in which we represent (left panel) the surface
brightness of NGC~1261 as a function of the radial distance. In this
plot, the points coming from our counts in each annulus have been
added to the data of Trager \etal (1995). The fourth annulus (which is
entirely outside the tidal radius given by Trager \etal) has been
divided into two sub-annuli.  It is clear that the surface brightness
is still decreasing outside $r_t$, and we can not use the LF from the
``external'' annulus to correct for foreground/background
contamination.  As a further evidence, in the right panel of the same
figure we compare the star counts observed in the region outside
$8^\prime$ with the counts predicted by Ratnatunga \& Bahcall (1985)
model.  Our counts are overabundant with respect to the expected
foreground up to a factor of five, well outside the uncertainty given
by Ratnatunga \& Bahcall for their model.  We conclude that all our
fields lie entirely inside the boundary of the cluster.  For this
reason, the background/foreground correction has been done using the
counts by Ratnatunga \& Bahcall (1985) based on the Galaxy model by
Bahcall \& Soneira (1984). The amount of this correction is, at most, 10\% 
of the original counts.

\begin{figure}
\resizebox{\hsize}{!}{\includegraphics{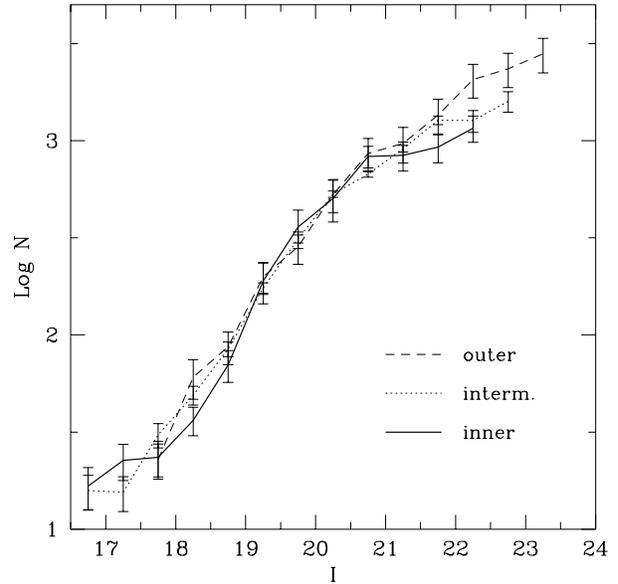}}
\caption{Total LFs for the three radial annuli. The counts have been
normalized in the magnitude interval 17.3$<I<$20.5.}
\label{fdl_tot}
\end{figure}

\begin{figure}
\resizebox{\hsize}{!}{\includegraphics{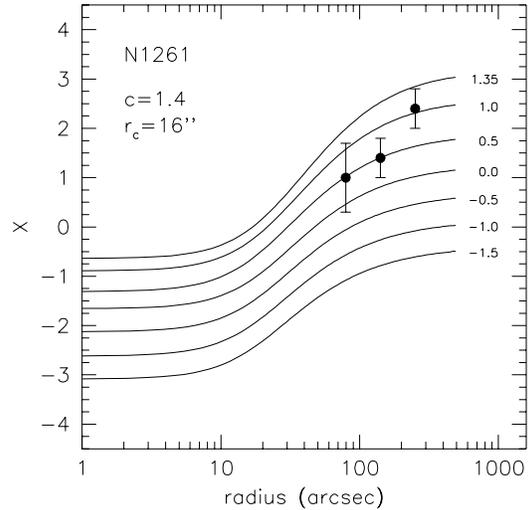}}
\caption{Multimass King--Michie models for a cluster with the 
concentration parameter of NGC~1261. The filled dots are the
observed values for the slope of the radial MFs.}
\label{pryor}
\end{figure}

\begin{figure*}
\resizebox{\hsize}{!}{\includegraphics{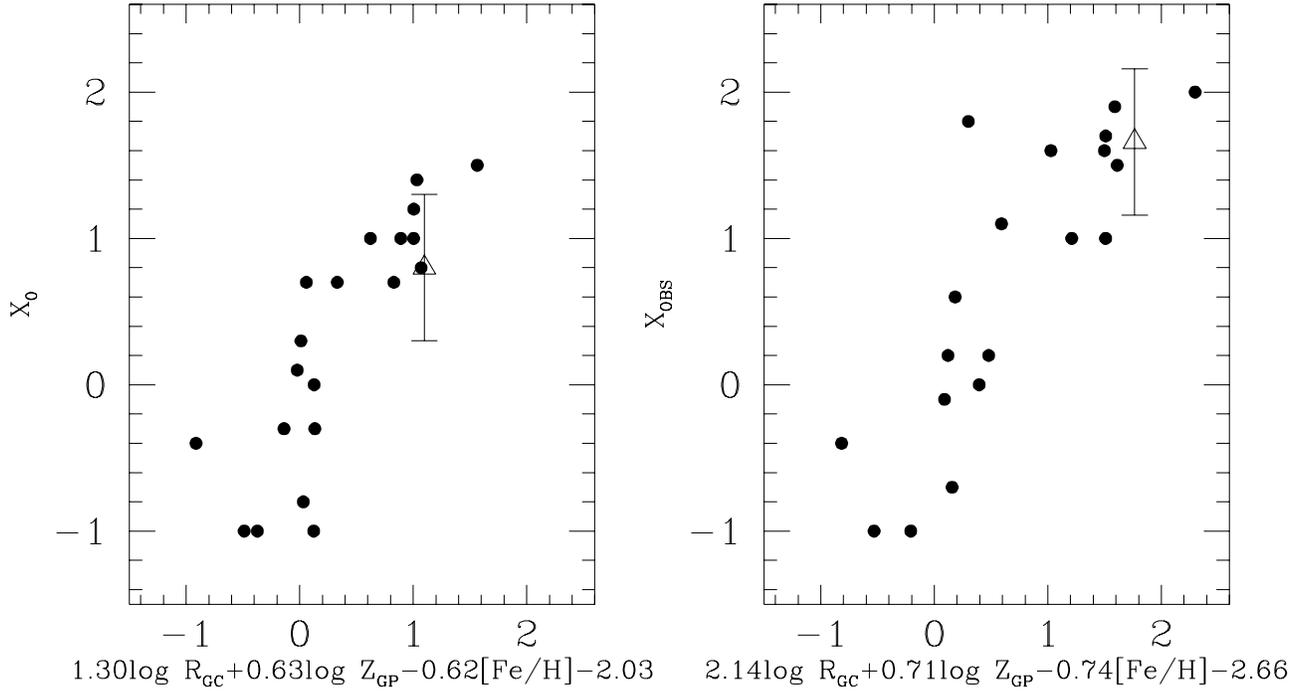}}
\caption{Trivariate relation by Djorgovski \etal (1993) with the point
relative to NGC~1261. The coefficients have been calculated 
using the new extended set of clusters with available MFs}
\label{dj_new}
\end{figure*}

No correction is made here for the possible contamination of the LF by
fainter, uresolved galaxies. However, a simple test was carried out to
insure that, would a correction be made, the conclusion of this work
would have been the same.  We examined the CMD obtained from two sets
of deep HST WFPC2 V and I images for the GCs M92 and M30, two clusters
whose projected position in the Galaxy is similar to that of
NGC~1261. We measured the number of foreground stars and galaxies for
these clusters counting all the objects that lie outside $\pm 5
\sigma$ from the cluster Main Sequence. We then scaled this counts to
1 arcmin$^2$ area and subtracted the number of object expected to be
stars, given by Ratnatunga \& Bahcall (1985). The number of remaining
objects is our best estimate of the number of galaxies per arcmin$^2$
that could contaminate our LF.  Due to their small area, this
correction does not affect at all the inner and intermediate annuli of
our LFs of NGC~1261.  It would change the last three magnitude bins of
the outer annuli of both fields. In particular the correction would
reach up to 5 counts per armin$^2$ in the last bin of the NTT frame
and 4 counts per arcmin$^2$ in the last bin of the 2.2m field.  The
corresponding MF slopes would be $x_0=0.6$ and $x_{OBS}=1.5$, well
inside the error bars we gave for our adopted values (cfr. section
\ref{sec3}). The slope for the outer MF would make the observed effect
of mass segregation more similar to what predicted by the model
(cf. Fig.\ref{pryor}).  However, this correction must be considered an
upper limit, because the completeness we have in the deep HST frames
is significantly larger than the completeness of our ground-based
frames for NGC~1261. It results quite difficult to estimate the
completeness of the background galaxy counts.  For this reason, and
for the very small statistic (hence large error) we have to deal with,
when counting foreground objects, we decided not to correct our
counts. The reader should simply keep in mind that our final values
are probably an upper limit for the outer MF slopes.  This slopes are
more likely to be comprised between those shown in Fig.\ref{dj_new} and
the two lower limit given above, when applying the foreground galaxy
correction.

\begin{figure}
\resizebox{\hsize}{!}{\includegraphics{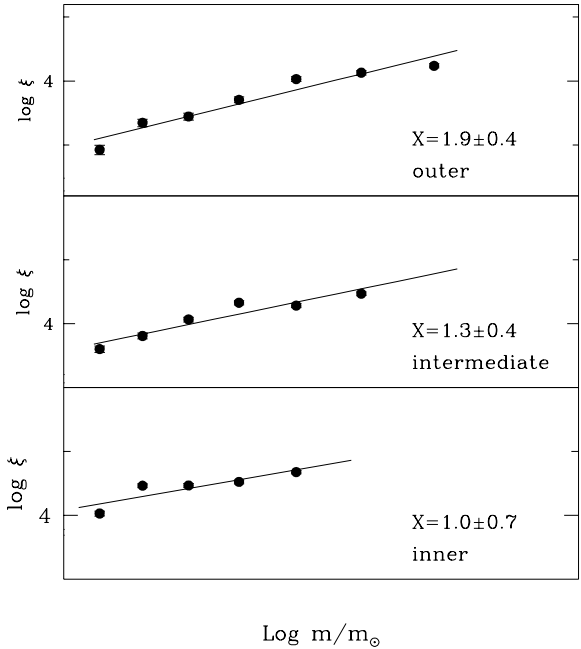}}
\caption{The mass functions of NGC~1261 in the three radial bins.
Note that the slope increases from the inner to the outer annulus as
a consequence of the mass segregation.}
\label{fdm_tot}
\end{figure}

The radial luminosity functions for each annulus of our two fields are
listed in Tables 1 and 2.  The counts are normalized per square
arcmin. Only counts with a completeness correction fraction $<2$ have
been used. Note that, where the dominant effect of the crowding is the
loss of stars, the corrected counts are larger than the raw ones,
whereas where the dominant effect is the migration of stellar
magnitudes (i.e.  in the brighter bins of the LF and/or in the outer
annuli) the corrected counts are sligtly smaller than the original
ones. This is due to the combination of two effects: the fact that the
migration of magnitudes moves the stars essentially toward brighter LF
bins, and the fact that this effect is larger for fainter stars.  The
errors of the incomplete LF represent the standard deviation of a
Poisson distribution, i.e. the square root of the number of stars; the
error in the complete counts were obtained combining them with the
errors in the inverse matrix. The outer annulus of the 2.2~m field is
the most noisy, because it contains a smaller number of stars.
Figure~\ref{fdl_rad} shows a comparison between the LFs of the
corresponding annuli of the two fields.  There is a good agreement
between the two fields, so the LFs used to derive the radial MFs have
been obtained by adding the LFs of the corresponding annuli in the two
fields. In the intermediate and outer annuli we excluded the last
magnitude bin of the NTT LF as there is not the corresponding bin in
the 2.2m data.  The final LFs for the three annuli are shown in
Fig.~\ref{fdl_tot}. Note how the LF becomes steeper and steeper moving
to the outer parts of the cluster. As discussed in the next Section,
this is what we expect as a consequence of mass segregation.


\section{Mass--Functions}
\label{sec3}

In order to convert the luminosity function $\Phi(M_I) \mbox{d} M_I$
into the mass function $\xi (m) \mbox{d} m$ we need a mass--luminosity
relation (MLR) appropriate for the Population II stars and for the
cluster metal abundance.  The most recent I--band MLR is from
Alexander \etal\ (1997).  For NGC~1261 we used the isochrones
corresponding to a metallicity [Fe/H]=--1.3 and we performed an
interpolation between the published isochrones relative to 10~Gyr and
20~Gyr in order to obtain an age of 15~Gyr (Alcaino \etal\ 1992).
According to the discussion in Section 1, we adopted a distance
modulus of $(m-M)=16.00$ and we neglected the absorption which is very
small, in view of the small reddening and of the fact that we observed
in the $I$--band.

The mass functions for the three radial annuli are shown in
Fig.~\ref{fdm_tot}. For consistency with the previous work, a straight
line has been fitted to the data points of both fields, in the mass
interval $0.5<m/m_\odot<0.8$. The slopes of these lines and their
errors, obtained by performing a weighted least square fit, are
written in each panel of the figure.

\subsection{Mass segregation correction}
The slope of the global MF of the stars in NGC~1261 can be obtained
after a correction for mass segregation.  We used isotropic multimass
King (1966) models following Pryor, Smith \& McClure (1986) and Pryor
\etal (1991).

For the multimass models we followed the same recipe and used the same
code of Pryor \etal\ (1991). Our only change was in the adopted MF,
which was assumed to be a power law and divided in 7 mass bins from
the turn-off mass down to $0.1$~M$_{\odot}$, plus a mass bin for the
remnant stars (in total 8 mass bins). The MF slope was varied in the
range $-1.5<x<1.35$, finding for each slope the model best fitting the
cluster radial density profile of Fig.~\ref{profilo}; then we
calculated the radial variation of $x$, due to mass segregation,
fitting the local MF in the model, in the same mass range of the
observed MF: $0.5<m/m_{\odot}<0.8$.  As it can be seen in
Fig.~\ref{pryor}, the MF slopes increase from the inner to the outer
bins, as expected (see also, Pryor \etal 1986).  Our three values of
the MF are consistent with the mass segregation predicted by the King
models and the global MF slope can be estimated to be $x_0=0.8\pm
0.5$.  We also determined the total MF, obtained by simply summing the
number of stars observed in each magnitude interval of each annulus.
This MF has a slope $x_{OBS}=1.7\pm 0.5$. It is not surprising that
this MF is steeper than the global one, as it refers to stars located
at $r>1\primip4$, i.e. $r>4.4r_c$, and not to the entire stellar
population in NGC 1261: this is a further evidence of the mass
segregation effects.

The structural parameters of our best fitting dynamical modelling of
NGC~1261 are different from the Trager et al values.: $r_c=16''$ 
and $c=1.4$ while Trager found $r_c=27''$ and $c=1.27$.  This is mainly
because adding our star counts to the Trager et al luminosity profile
we gave a different weight to the outer part of the cluster profile.
In any case this difference of structural parameters do not change our
previous results on the global value of $x$ since our models are
directly best fitted to the data and not simply calculated from the
structural parameters.

%

\section{Discussion}
The values of $x_0$ and $x_{OBS}$ have been added to the figures with
the correlation between the MF slopes of the other 20 clusters and
their position and metallicity as proposed by Djorgovski \etal (1993).
As shown in Fig.~\ref{dj_new}, taking into account the uncertainties
in the slope, also the MF of NGC~1261 follows the correlation proposed
by Djorgovski \etal (1993). In order to have a quantitative estimate
of how well the distribution of Fig.\ref{dj_new} is fitted by a
straight line, we calculated the correlation factors, that should be
very close to 1 if the distribution is actually a straight line. For
the points in the two panels of Fig.~\ref{dj_new} we obtained:
\begin{displaymath}
r(x_0)=0.835~~~~\mbox{and}~~~~r(x_{OBS})=0.843 .
\end{displaymath}

We calculated the new coefficients for the trivariate formula,
including the points for NGC~1261, NGC~3201 (Brewer \etal 1993) and
NGC~1851 (Saviane \etal 1997) and M55 (Zaggia \etal 1997) using the
same method described in Djorgovski \etal. The new data suggest that
the best trivariate relations are:

\begin{displaymath}
x_{obs}=(2.14\pm 0.85)\log R_{GC}+(0.71\pm 0.36)\log Z_{GP} 
-(0.74\pm 0.30)\mbox{[Fe/H]}-2.66
\end{displaymath}

for the observed MF slope, and:

\[
x_0=(1.30\pm 0.66)\log R_{GC}+(0.63\pm 0.29)\log Z_{GP} 
-(0.62\pm0.24)\mbox{[Fe/H]}-2.03
\]

for the mass segregation corrected MF slope.

\begin{acknowledgements}
We are grateful to C. Pryor for providing us the dynamical code for
the mass segregation correction.
\end{acknowledgements}


%



\end{document}